\documentclass[aps,prl,reprint,preprintnumbers,superscriptaddress,nofootinbib]{revtex4-1}
\usepackage{graphicx} 
\usepackage{dcolumn} 
\usepackage{bm} 
\usepackage[pdfencoding=auto, psdextra]{hyperref} 
\usepackage{amsmath}

\usepackage{epstopdf}
\usepackage{epsfig}
\usepackage{graphics}
\usepackage{xspace}
\usepackage{slashed}
\usepackage{xcolor}
\usepackage{longtable}
\usepackage[small]{subfigure}
\usepackage[normalem]{ulem}
\usepackage{multirow}

\newcommand{\mbar}{\overline{m}}

\newcommand{\MSb}{\overline{\mathrm{MS}}}

\newcommand{\cO}{\mathcal{O}}

\newcommand{\beq}{\begin{equation}}
\newcommand{\eeq}{\end{equation}}

\begin{document}

\title{Precise \texorpdfstring{\boldmath $\alpha_s$}{alphas} determination from charmonium sum rules}

\author{Diogo Boito}
\affiliation{Instituto de F\'isica de S\~ao Carlos, Universidade de
             S\~ao Paulo, CP 369, 13560-970, S\~ao Carlos, SP, Brazil}

\author{Vicent Mateu}
\affiliation{Departamento de F\'isica Fundamental e IUFFyM,\\Universidad de Salamanca, E-37008 Salamanca, Spain}
\affiliation{Instituto de F\'isica Te\'orica UAM-CSIC, E-28049 Madrid, Spain}

\date{\today}

\begin{abstract}
The strong coupling, $\alpha_s$, governs perturbative Quantum Chromodynamics (QCD) and is one of the free parameters of the Standard Model.
We introduce a new method that allows a precise extraction of  $\alpha_s(m_Z)$  
from dimensionless ratios of roots of moments of the charm-quark
vector correlator. The ratios we use in our analysis have a rather weak logarithmic quark-mass dependence, starting at $\cO(\alpha_s^2)$, and can be
obtained from experimental data with good precision, since they benefit from  positive correlations among the individual  experimentally determined
moments. We perform a careful and conservative error analysis with special emphasis on uncertainties related to the truncation of perturbation theory,
treating the renormalization scales such as to ensure order-by-order convergence. Our final result, with expressions at $\cO(\alpha_s^3)$, is
$\alpha_s(m_Z)=0.1168\pm 0.0019$.
\end{abstract}

\pacs{QCD,pQCD}

\preprint{IFT-UAM/CSIC-19-166}

\maketitle

The  strong coupling, $\alpha_s$, is one of the fundamental parameters of the Standard Model (SM). It is the expansion parameter governing
perturbative QCD expansions and  its value cannot be  predicted by theory; the extraction of the coupling always requires the comparison of
quantities calculated in state-of-the-art QCD with experimental --- or lattice --- data. Apart from its prominent role in precision QCD, flavour
physics, and the calculation of hadronic properties, a good control over the value of $\alpha_s$ is key  
for LHC physics, in particular to have a reliable determination of parton distribution functions, which are largely correlated with the strong coupling.
In forthcoming $e^+e^-$ colliders, with dedicated Higgs and top-quark precision measurement programs, $\alpha_s$ will remain a crucial input.
Additionally, the values of $\alpha_s$ and of the top-quark mass are behind the fate of the SM vacuum~\cite{Degrassi:2012ry}.

Significant progress has been made in the past few years to extract $\alpha_s$ with good precision, which requires effort both in experimental
measurements or lattice simulations, as well as in theoretical computations, in order to reach higher levels of accuracy which depend, in particular, on
calculations at higher loop order. Extractions based on lattice data, especially, have improved considerably in the recent past. However,
several tensions still remain, which has led the Particle Data Group to almost double the uncertainty on its recommended  $\alpha_s(m_Z)$ world
average since the 2016 edition~\cite{Tanabashi:2018oca,Patrignani:2016xqp}. It remains, therefore,
very important to find reliable observables to extract the strong coupling, in which both theory and experiment are under very good control. In this
paper we describe for the first time the use of ratios of roots of moments of the charm-quark vector correlator in precise extractions of $\alpha_s$.

One of the standard observables in QCD is the total cross section for $e^+e^- \to {\rm hadrons}$ and the associated $R_{q\bar q}(s)$ ratio defined as
\begin{equation}\label{eq:Rqq}
R_{q\bar{q}}(s) =\! \frac{3s}{4\pi\alpha^2}\sigma_{e^+e^-\to\, q\bar{q}\,+X}(s) \simeq
\dfrac{\sigma_{e^+e^-\to\, q\bar{q}\,+X}(s)}{\sigma_{e^+e^-\to\,\mu^+\mu^-}(s)},
\end{equation}
where $q=c,b$ is the quark species, $\alpha$ the fine-structure constant, $\sqrt{s}$ the $e^+e^-$ center-of-mass energy, and the right-hand side is
exact when the denominator is calculated  in the limit of massless muons and at leading order in $\alpha$.\footnote{Even though, strictly speaking, the
process is mediated both by a photon and a $Z$ boson, at the energies relevant for the moments it is overwhelmingly dominated by the former,
which moreover is a vector current. An estimate of the (small) axial-vector contribution can be found in Table~8 of Ref.~\cite{Dehnadi:2011gc}.} Integrated
moments of $R_{q\bar q}(s)$ play a prominent
role, since they make use of data in broad energy regions, as opposed to considering the observable locally, which can significantly improve their
experimental precision and the reliability of their theoretical description. These integrated moments can also be, in many cases, rigorously calculated
in perturbation theory.  In this work, the inverse moments of $R_{c\bar c}(s)$ defined as
\begin{equation}
\label{eq:momentdefvector2}
M_c^{(n)} = \int_{s_0}^{\infty}\!\dfrac{{\rm d}s}{s^{n+1}}R_{c\bar{c}}(s),
\end{equation}
are specially important,  where $s_0$ must be smaller than the squared mass of the first $c\bar c$ narrow resonance, the $J/\psi$. They have
been, so far, mainly used in the precise extraction of the $c$- and $b$-quark masses from data. In the present work, for reasons that will become
clear soon, we are interested in dimensionless ratios of roots of moments~$M_c^{(n)}$,
\begin{equation}\label{eq:ratMM}
R_c^{V,n}\equiv \frac{\big(M_c^{(n)}\big)^\frac{1}{n}}{\big(M_c^{(n+1)}\big)^\frac{1}{n+1}} ,
\end{equation}
where $V$ refers to the fact that the moments are related to the vector charm-quark current correlator. Analogous ratios of moments have originally
been introduced in the context of the pseudo-scalar charm correlator for which only lattice data is available~\cite{Maezawa:2016vgv}.
As we will show, the ratios $R_c^{V,n}$ that we introduce here are particularly suitable for $\alpha_s$ extractions: for $1\leq n \leq 3$ they are known
up to $\cO(\alpha_s^3)$, have a very weak dependence on the $c$-quark mass, and can be accurately determined using the experimental values for the
masses and partial widths of narrow resonances, supplemented with continuous data for $R_{c\bar c}(s)$.

Let us start by discussing the perturbative expansion for $M_c^{(n)}$ and the ratios $R_c^{V,n}$. Using analyticity and unitarity, the moments
$M_c^{(n)}$ can be related to derivatives of the vector charm-quark current correlator. The theoretical counterpart to
Eq.~\eqref{eq:momentdefvector2} reads~\cite{Shifman:1978bx,Shifman:1978by}
\begin{equation}
\label{eq:MqVTh}
M_c^{(n)} =\dfrac{12\pi^2 Q_c^2}{n!}\,\dfrac{{\rm d}^n}{{\rm d}s^n}\Pi_c(s)\Big|_{s=0},
\end{equation}
where $Q_c$ is the charm-quark electric charge and the correlator is formed from the charm vector currents as
\begin{equation}
\!\big(g^{\mu\nu}s-p^{\mu}p^{\nu}\big)\Pi_c(s) =\! - i\!\!\int\!\!{\rm d}x\, e^{i\,p\cdot x}
 \langle 0|T\,j_c^{\mu}(x)j_c^{\nu}(0)|0\rangle,\!
\end{equation}
with $j_c^{\mu}(x) = \bar{c}(x)\gamma^\mu c(x)$. The Taylor coefficients of the $\Pi_c(s)$ expansion in powers of $s$  around $s=0$, that participate
in Eq.~\eqref{eq:MqVTh}, can be accurately calculated in perturbation theory with the typical short-distance scale given by
$\sim m_c/n > \Lambda_{\rm QCD}$ (restricting $n$ to small values). In full generality, the perturbative expansion of $M_c^{(n)}$ is written in terms of
two renormalization scales, $\mu_\alpha$ and $\mu_m$, at which the strong coupling and the quark-mass are respectively evaluated, as first noticed in
Ref.~\cite{Dehnadi:2011gc}:
\begin{align}\label{eq:MoMmm}
&M^{(n)}_c = \frac{1}{[2\,\mbar_c(\mu_m)]^{2n}} \sum_{i=0} \Biggl[\frac{\alpha_s^{(n_f)}(\mu_\alpha)}{\pi}\Biggr]^i\\
&\times \sum_{a=0}^{i} \sum_{b=0}^{[i-1]} \, c^{(n)}_{i,a,b}(n_f) \ln^a\!\biggl(\frac{\mu_m}{\mbar_c(\mu_m)}\biggr)
\ln^b\!\biggl(\frac{\mu_\alpha}{\mbar_c(\mu_m)} \biggr),\nonumber
\end{align}
with $[i-1]\equiv {\rm Max}(i-1,0)$, and $n_f=4$. The running mass $\mbar_c(\mu_m)$ and coupling $\alpha_s(\mu_\alpha)$ are calculated in the $\MSb$
scheme with the five-loop QCD $\gamma$ and $\beta$ functions, respectively~\cite{Baikov:2014qja,Luthe:2016xec,Luthe:2017ttg}. Likewise, we use the
four-loop matching condition~\cite{Chetyrkin:1997un,Chetyrkin:2005ia,Schroder:2005hy} to relate $\alpha_s$ in the four- and five-flavour schemes. (We
will often omit the explicit $n_f$ dependence in $\alpha_s$ and $c^{(n)}_{i,a,b}$.) The leading logarithm in $M_c^{(n)}$ appears at order
$\alpha_s$. Setting the two scales in Eq.~\eqref{eq:MoMmm} to the common value $\mu_\alpha=\mu_m=\mbar_c(\mbar_c)$ the logarithms are
resummed and the expansion of $M_c^{(n)}$, in this particular case, exposes the independent coefficients $c^{(n)}_{i,0,0}$ which must be calculated
in perturbation theory.
Thanks to a tremendous computational effort, the coefficients $c^{(n)}_{i,0,0}$ have been calculated (analytically) for $n=1,\,2,\, 3$ and
$4$~\cite{Maier:2008he, Maier:2009fz,Maier:2017ypu} up to  order $\alpha_s^3$ [\,four loops, or next-to-next-to-next-to-leading order (N$^3$LO)\,].
For $n>4$ only estimates are available at this order \cite{Hoang:2008qy, Kiyo:2009gb, Greynat:2010kx, Greynat:2011zp}. The logarithms of Eq.~\eqref{eq:MoMmm}
with the respective  coefficients can be generated with the use of renormalization group equations. Numerical values of the coefficients $c^{(n)}_{i,a,b}$ can be
found in Ref.~\cite{Dehnadi:2015fra}. The dependence of $M_c^{(n)}$ on $\mbar_c$ through the prefactor  
makes these moments ideal for the extraction of the charm-quark mass.

The ratios we are interested in, given in Eq.~\eqref{eq:ratMM}, are  constructed in such a way as to cancel the
mass dependence of the prefactor in Eq.~\eqref{eq:MoMmm}. Their fixed-order perturbative expansion reads
\begin{align}\label{eq:R2scales}
R^{V,n}_c &=
\sum_{i=0} \bigg[\frac{\alpha_s(\mu_\alpha)}{\pi}\bigg]^i  \\  & \times\sum_{k=0}^{[i-1]}\sum_{j=0}^{[i-2]} r^{(n)}_{i,j,k}
\ln^j\biggl(\frac{\mu_m}{\mbar_c(\mu_m)}\biggr)
\ln^k\biggl(\frac{\mu_\alpha}{\mbar_c(\mu_m)}\biggr),\nonumber
\end{align}
where now the first logarithm, which brings the dependence on $\mbar_c$, appears only at $\alpha_s^2$. The ratios $R^{V,n}_c$ are, therefore, almost
insensitive to the quark mass. The coefficients $r_{i,j,k}^{(n)}$ can be obtained from $c^{(n)}_{i,0,0}$ upon re-expansion of $R_c^{V,n}$ in $\alpha_s$
and the use of renormalization group equations. For instance, for  $R_c^{V,2}$ at N$^3$LO one finds
\begin{align}\label{eq:RthNum}
&R_c^{V,2} = 1.0449\big[1+0.57448\,a_s \nonumber \\
&+\left(0.32576 + 2.3937\,L_{\alpha}\right)a_s^2 \\
&-\left(2.1093+ 4.7873 L_m -6.4009 L_\alpha - 9.9736 L_\alpha^2\right)a_s^3\big]\nonumber,
\end{align}
where here $a_s=\alpha_s(\mu_\alpha)/\pi$, $L_\alpha = \ln[\,\mu_\alpha/\mbar_c(\mu_m)\,]$  and  $L_m = \ln[\,\mu_m/\mbar_c(\mu_m)\,]$. The total
$\alpha_s$ correction to $R_c^{V,1}$ is about $12.5\%$, $7.2\%$ for $R_c^{V,2}$, and $5.2\%$ for $R_c^{V,3}$.
The perturbative contribution to $R_c^{V,n}$ is the first term in its Operator Product Expansion (corresponding to the identity operator). The leading
non-perturbative correction stems from the gluon condensate and is known to $\cO(\alpha_s)$~\cite{Broadhurst:1994qj}. This correction is  small,
but nevertheless included in our analysis even though our results are largely dominated by perturbative QCD.

Alternatively, one could consider not re-expanding in $\alpha_s$ the ratios defined in Eq.~\eqref{eq:ratMM}. In principle, one could even take different
renormalization scales in the numerator and denominator. Even though the pole-mass ambiguity cancels individually in each moment, subleading
renormalons exist and their effect might be softened by taking the same renormalization scale and re-expading the ratios. Furthermore, the physics
of $R_c^{V,n}$ is different from the one of each individual moment, and as such they should be considered as  observables in their own right, therefore
with their own series expansion in terms of a single $\alpha_s(\mu_\alpha)$.

We turn now to the experimental determination of the ratios $R_c^{V,n}$. Our results are based on the obtention of the inverse moments $M_c^{(n)}$
performed in Ref.~\cite{Dehnadi:2011gc} and discussed in detail in that work. It combines the contribution from the narrow $J/\psi$ and $\psi^\prime$
resonances,  the available threshold data from Refs.~\cite{Bai:1999pk,Bai:2001ct,Ablikim:2004ck,Ablikim:2006aj,Ablikim:2006mb,:2009jsa,Osterheld:1986hw,
Edwards:1990pc,Ammar:1997sk,Besson:1984bd,:2007qwa,CroninHennessy:2008yi,Blinov:1993fw,Criegee:1981qx,Abrams:1979cx}, and a remaining
contribution modeled with perturbative QCD for $s>10.538$\,GeV where no data is available (the so-called continuum contribution). One also subtracts
from the data a non-charm background from $u$, $d$, and $s$ quarks, as well as a contribution from secondary charm production which is not included in
the theory. (The small singlet contribution has been estimated and can be neglected~\cite{Kuhn:2007vp}.) The continuum contribution as well as the $uds$
background, which are implemented at the $R$-ratio level, use perturbative QCD expressions. Here, since we aim at a precise extraction of $\alpha_s$,
we cannot fix its value in these contributions. We
have, therefore, adapted the extraction of the moments $M_c^{(n)}$ from Ref.~\cite{Dehnadi:2015fra} in order to obtain $R_c^{V,n}$ as a function of the
$\alpha_s$ value used in the continuum and the background. It turns out that the dependence with $\alpha_s$, for values not too far from the world average,
is  highly linear, which facilitates the task of obtaining parametrized expressions for the ratios $R_c^{V,n}$. In terms of
$\Delta_\alpha=\alpha_s^{(n_f=5)}(m_Z)-0.1181$, the three ratios we exploit here read
\begin{align}\label{eq:ExpRmom}
R_c^{V,1}&= (1.770- 0.705\,\Delta_\alpha) \pm 0.017 \nonumber,\\
R_c^{V,2}&= (1.1173 -0.1330\,\Delta_\alpha)\pm 0.0022 , \\
R_c^{V,3}&= (1.03535 - 0.04376\,\Delta_\alpha)\pm 0.00084 .\nonumber
\end{align}
The associated errors are dominated by data and are fairly small. The smallness of the uncertainties is in part due to the strong positive correlations
between the consecutive moments $M_c^{(n)}$ which,  in the error propagation, lead to a very small uncertainty in the ratios. (For example, moments
$M_c^{(2)}$ and $M_c^{(3)}$ are $97.6\%$ correlated.) The relative errors in the ratios are of only $0.98\%$, $0.22\%$, $0.10\%$ for $R_c^{V,1}$,
$R_c^{V,2}$, and $R_c^{V,3}$, respectively.

\begin{figure}[!t]
\includegraphics[width=0.75\columnwidth]{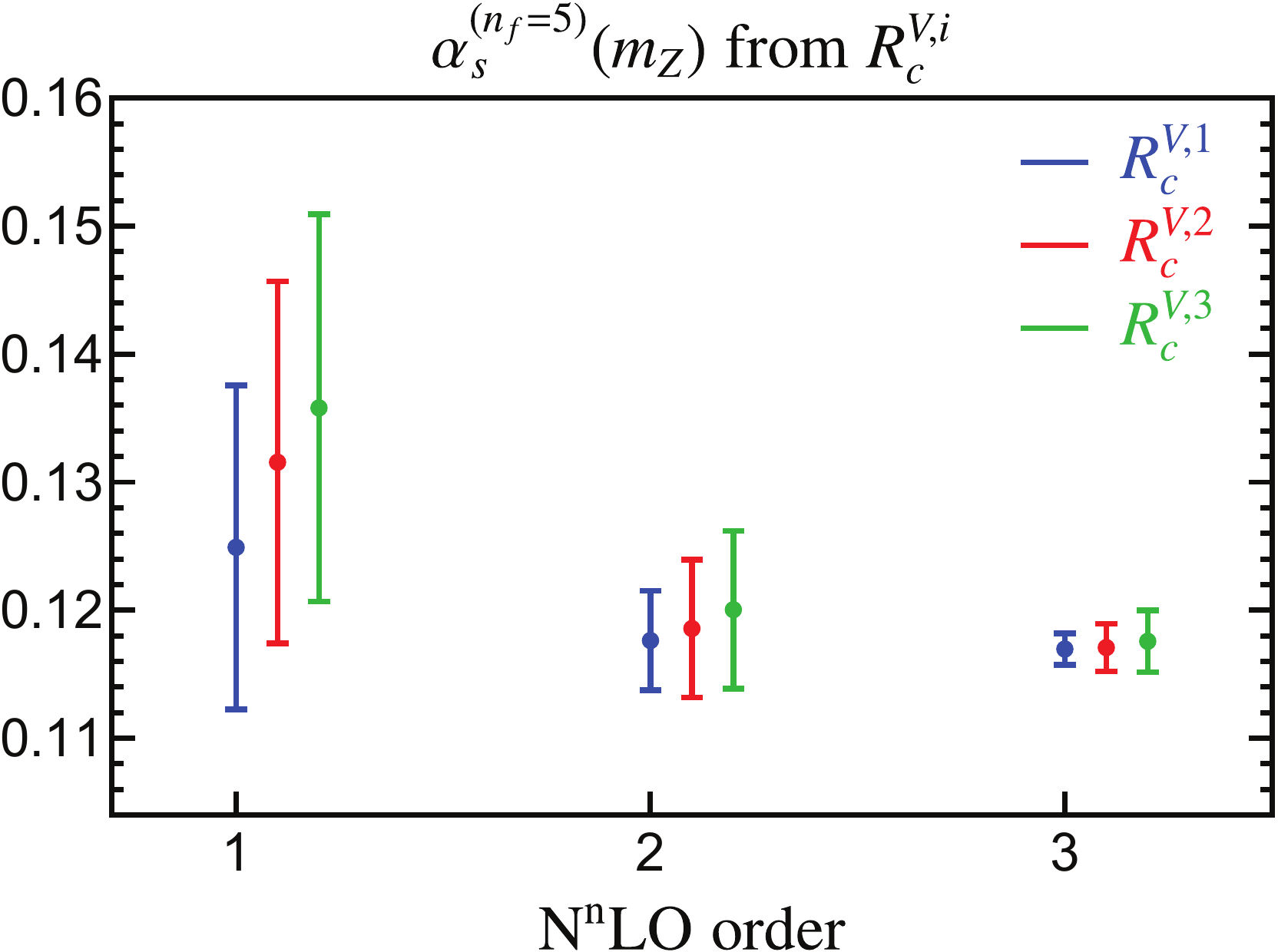}
\caption{$\alpha_s$ values extracted order by order in perturbation theory from the  ratios $R_c^{V,n}$ of Eq.~\eqref{eq:ExpRmom}. Only
perturbative uncertainties are displayed.\label{Fig1}}
\end{figure}

The determination of $\alpha_s$ is done by equating the experimental results of Eq.~\eqref{eq:ExpRmom} to the respective expansions of the type of
Eq.~\eqref{eq:RthNum}, numerically solving for $\alpha_s$. We turn now to a discussion of the results we obtain from this analysis. Sound results
require a careful --- and conservative --- study of the associated uncertainties, in particular those that stem from the truncation of the perturbative
series. It has been shown that in quark-mass extractions from $M_c^{(n)}$, a reliable error estimate requires the independent variation of the two
scales $\mu_m$ and $\mu_\alpha$~\cite{Dehnadi:2015fra}. To be fully conservative, even though here the dependence on $\mu_m$ is weaker than in
the case of $M_c^{(n)}$, we vary both scales in the interval \mbox{$\mbar_c \leq \mu_\alpha,\mu_m \le \mu_{\rm max}$}, with $\mu_{\rm max}=4\,$GeV,
and apply the constraint $1/\xi\leq (\mu_\alpha/\mu_m)\leq \xi$ with the canonical choice $\xi=2$ (the dependence on the value of $\xi$ will be
discussed below).\footnote{We have carefully investigated the convergence of the perturbative expansion with an adapted Cauchy test suggested in
Ref.~\cite{Dehnadi:2015fra} and conclude that the use of the restriction $1/\xi\leq (\mu_\alpha/\mu_m)\leq \xi$ is sound in our case.} The scale variation
we adopt is much more conservative than that used in many related works, where one often sets $\mu_m=\mu_\alpha$ (or $\xi=1$). For the charm
mass we adopt $\mbar_c=1.28(2)$\,GeV.
With this setup we have created grids
with $3025$ points of $\mu_m$ and
$\mu_\alpha$ and the respective $\alpha_s$ values for each ratio $R_c^{V,n}$ (with $n=1,2,$ and $3$), order by order in the perturbative expansion.
First, we check the convergence of the $\alpha_s$ extractions at each order in perturbation theory from the results obtained in the grids, neglecting
charm-mass, experimental, and non-perturbative uncertainties. Therefore the spread in values due to scale variation directly measures the perturbative
error. The results are shown in Fig.~\ref{Fig1} for the three ratios we consider. One clearly sees a nice convergence for all the moments,
which indicates that the perturbative uncertainties are under control.

\begin{figure}[!t]
\includegraphics[width=0.7\columnwidth]{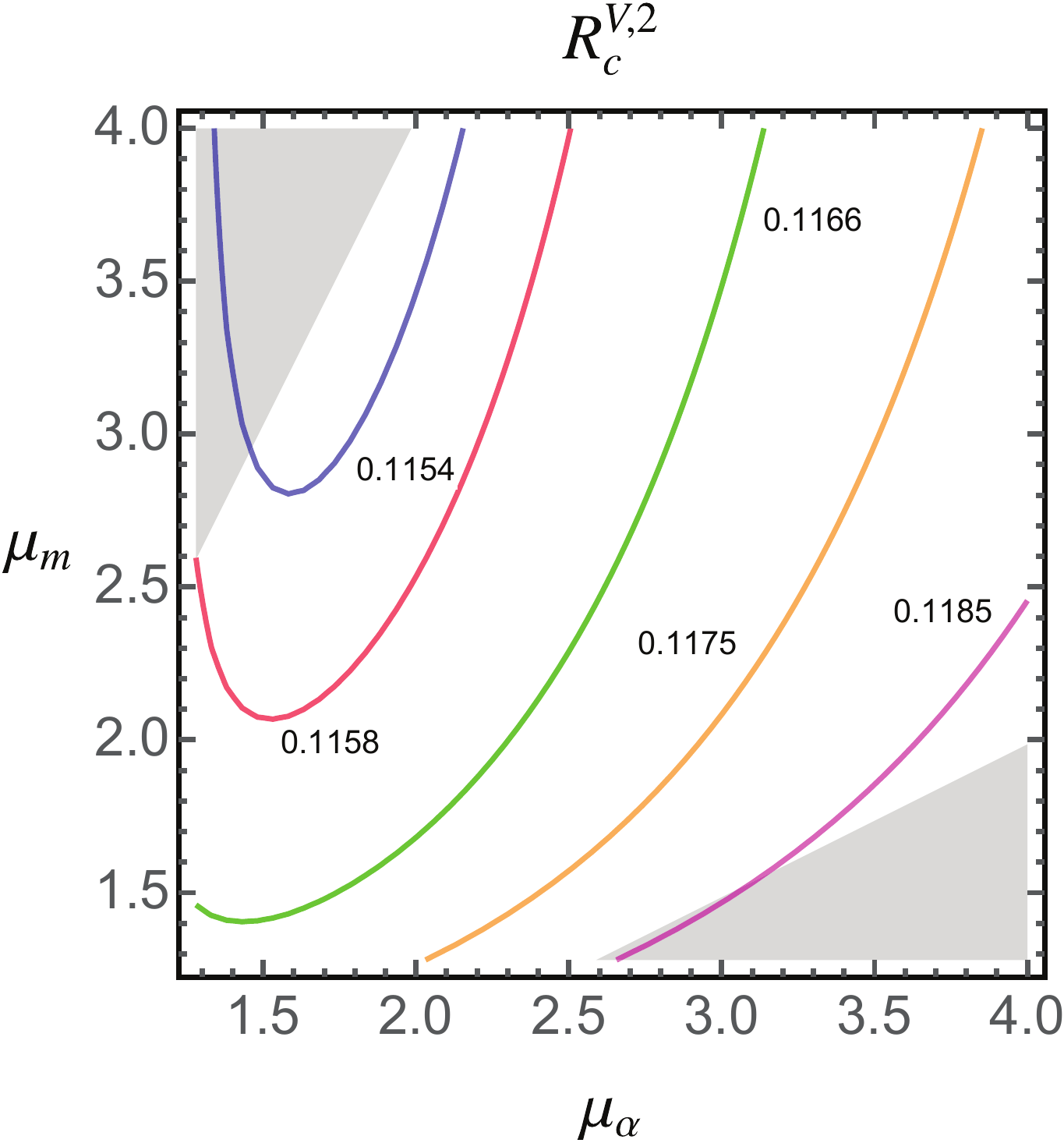}
\caption{Results for $\alpha_s$ from $R_c^{V,2}$ at  $\cO(\alpha_s^3)$ in the  $\mu_\alpha \times \mu_m$ plane. Shaded areas are excluded from
our  analysis (see text). \label{Fig2}}
\end{figure}

We continue the investigation of perturbative incertitudes by analyzing the $\alpha_s$ grids with two-dimensional contour plots at N$^3$LO. In
Fig.~\ref{Fig2} we show the result of such a scan  in the case of $R_c^{V,2}$. What one sees from this plot is that a correlated scale variation with
$\mu_\alpha=\mu_m$, along the diagonal of the plot, would lead to a seriously underestimated theory uncertainty. The consequences of a correlated
scale variation would be less dramatic for $n=1$   but the results of Fig.~\ref{Fig2} demonstrate, visually, the need for the independent
scale variation. Finally, to examine systematically the consequences of less (and more) conservative scale variations, we vary the value of $\xi$
between $\xi=1$, which corresponds to $\mu_\alpha=\mu_m$, and $\xi=3$, that imposes almost no constraint within our intervals.  For $\xi=1$ we
find that the perturbative uncertainties would be underestimated by factors of $3$ ($n=1$), $2$ ($n=2$), and $1.5$ ($n=3$) compared to our canonical
choice ($\xi=2$). On the other hand, adopting an even more conservative choice with $\xi=3$, would lead to increases in the errors between $30$\% and
$60\%$, which shows that our canonical choice is
sufficient for a conservative error estimate. The central values of $\alpha_s$ are rather stable with the choice of $\xi$ and the
variations are below the percent level for $1\leq\xi\leq3$.

With the perturbative uncertainties under good control, we are in a position to extract the final values of our analysis. To study the other sources of
uncertainties we created additional $\alpha_s$ grids in the $\mu_m\times \mu_\alpha$ plane varying within one sigma the experimental value of
$R_c^{V,n}$,  the charm-quark mass, and also adding and removing twice the gluon-condensate contribution (as an estimate of non-perturbative
uncertainties). We find, through the analysis of these grids,
\begin{align*}
  \alpha_s(m_Z) &= 0.1168(10)_{\rm pt}(28)_{\rm exp}(6)_{\rm np}=0.1168(30)\,\,[ R_c^{V,1}],  \\
  \alpha_s(m_Z) &= 0.1168(15)_{\rm pt}(9)_{\rm exp}(7)_{\rm np}=0.1168(19)\,\,   [R_c^{V,2}],  \\
  \alpha_s(m_Z) &= 0.1173(20)_{\rm pt}(5)_{\rm exp}(6)_{\rm np}=0.1173(22)\,\,  [R_c^{V,3}],
\end{align*}
where the first error is due to the truncation of perturbation theory, obtained from the the spread of values arising from the independent scale
variation with $\xi=2$, the second comes from the experimental errors given in Eq.~\eqref{eq:ExpRmom}, and the third is due to non-perturbative
contributions. Perturbative errors grow with $n$ while experimental errors become smaller. The error  for the result
with $n=1$ is largely dominated by experiment, while for $n=2$ and $n=3$ the perturbative error dominates. In all cases the
uncertainty associated with the charm-quark mass is $0.0003$ and does not contribute to the final error. The non-perturbative error is always
subleading, but gives a small contribution to the total error for  $n=2$.

\begin{figure}[!t]
\includegraphics[width=0.9\columnwidth]{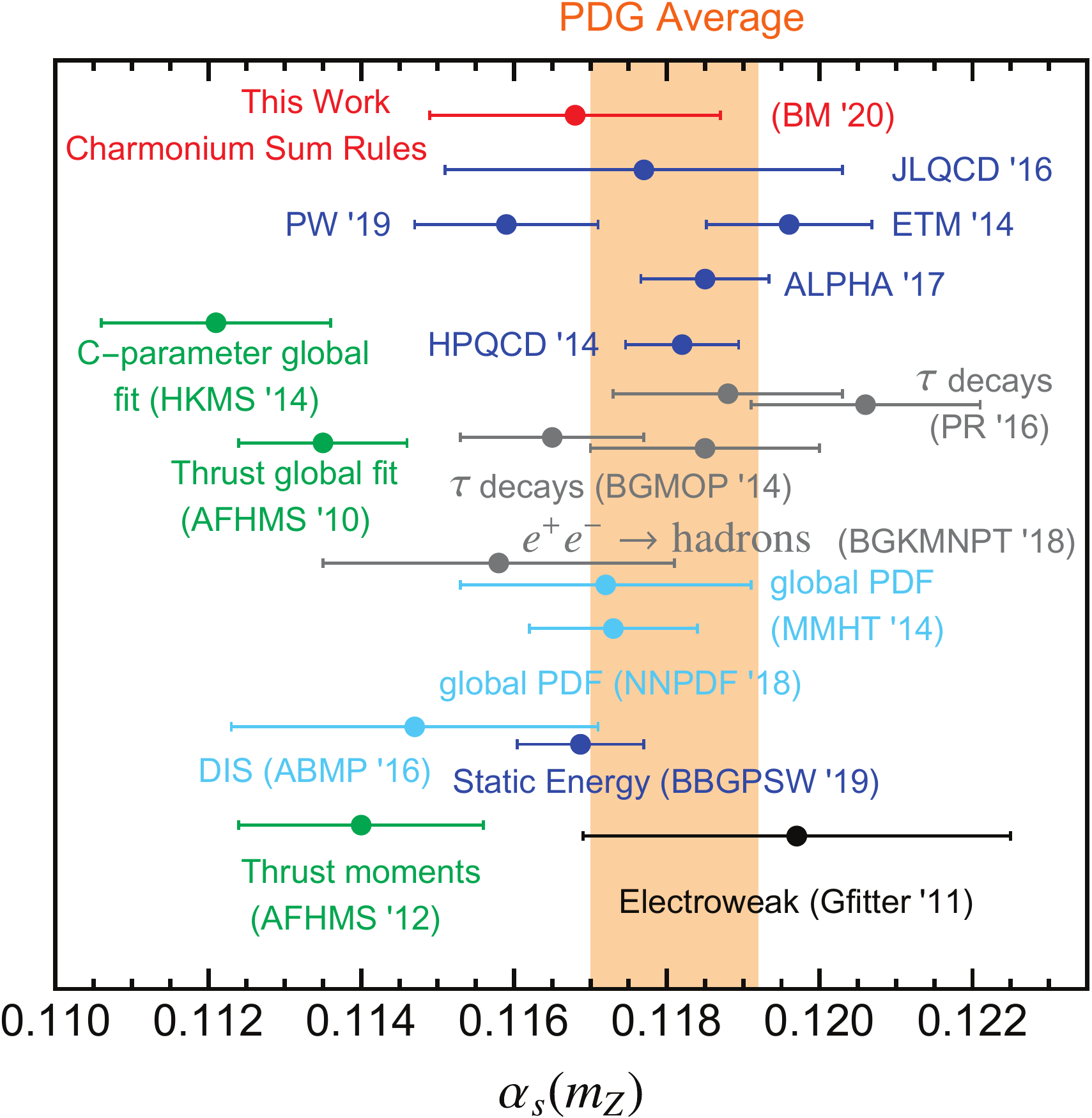}
\caption{Comparison of our determination of $\alpha_s^{(n_f=5)}(m_Z)$ (top, in red) with a few recent determinations. Event-shape analyses at
N$^3$LL$^\prime$ + $\mathcal{O}(\alpha_s^3)$: thrust and \mbox{C-parameter} (green)~\cite{Abbate:2010xh,Abbate:2012jh,Hoang:2015hka};
lattice QCD~\cite{HPQCD:2014aca,Petreczky:2019ozv,Bruno:2017gxd,Blossier:2013ioa,Nakayama:2016atf} and static energy
potential~\cite{Bazavov:2019qoo} (in dark blue); Electroweak precision observables fits~\cite{Flacher:2008zq} (black); Deep Inelastic
Scattering~\cite{Alekhin:2017kpj} and global PDF fits \cite{Ball:2018iqk,Harland-Lang:2015nxa} (light blue); and hadronic $\tau$
decays~\cite{Boito:2014sta,Pich:2016bdg} and $e^+e^-\to$ hadrons~\cite{Boito:2018yvl} (gray). The current world average~\cite{Patrignani:2016xqp}
is shown as an orange band.}\label{Fig3}
\end{figure}

The final results for $\alpha_s$ are correlated since they are based on ratios of moments obtained from the same data sets. This disfavors averaging
the results obtained from the different ratios $R_c^{V,n}$. Instead, we quote as our final value the one obtained from the ratio $R_c^{V,2}$ for the
following reasons:
a)~the experimental uncertainty, in the case of the extraction from $R_c^{V,1}$, is significantly larger, which
makes the final error much less competitive; b)~the extraction from $R_c^{V,3}$, on the other hand, relies on $M_c^{(4)}$, which may have a too large
value of $n$ and correspondingly a smaller effective scale --- a fact that is also responsible for the  larger perturbative uncertainty. The
most reliable result is therefore the one from $R_c^{V,2}$ which yields our final value
\begin{equation}\label{eq:final}
  \alpha_s^{(n_f=5)}(m_Z) =0.1168\pm 0.0019.
\end{equation}
Our result is fully compatible with the present world average [\,$0.1181(11)$\,]~\cite{Tanabashi:2018oca} although the uncertainty is larger. Our
determination has a very conservative error estimate: with a correlated scale variation the uncertainty would be reduced to $0.0013$, not much
larger than the world average.
Comparison with other works in the literature~\cite{Petreczky:2019ozv,HPQCD:2014aca,Blossier:2013ioa} show that our perturbative error is also more conservative than what is obtained from
estimates of higher-order contributions (as opposed to scale variations). Our treatment of the experimental moments is also completely unbiased,
since we do not fix $\alpha_s$ to compute the perturbative contribution, but keep it as a free parameter. Using  experimental moments with $\alpha_s$ fixed to the world average in the perturbative contribution would lead to even smaller errors and  central values that change by an amount an order of magnitude smaller than the total uncertainty. Our procedure is, again, the most conservative alternative.
Our result is compared with other selected recent extractions of $\alpha_s$ in Fig.~\ref{Fig3}.

The present analysis can be extended in a number of directions.  First, it can directly be applied to the vector moments of the bottom-quark current.
Our preliminary
results show that the errors on $\alpha_s$ in this case are not as competitive as the ones from the charm. One can also apply our more conservative
treatment of perturbative uncertainties to analyze pseudo-scalar current moments obtained on the lattice. Our results on these additional analyses will
be presented elsewhere, together with further details on the results from the charm vector-current analysis. On the theory side, one could also investigate
alternative ways of organizing the perturbative expansion, such as using different powers of $R_c^{V,n}$ (re-expanded in $\alpha_s$) or linearized
iterative solutions (in the spirit of~\cite{Dehnadi:2015fra}). Additionally, the cancellation of the renormalon associated with the pole mass when taking
ratios allows for an analysis that employs directly the pole mass in the logarithms. One could also consider fits using all available information (including
correlations) in order to extract  $\alpha_s$  and the quark-masses in a self-consistent way.  We plan to carry out these analyses in the near future.

\begin{acknowledgments}
{\bf Acknowledgments:}
This work was supported in part by the SPRINT project funded by the S\~ao Paulo Research Foundation (FAPESP) and the University of Salamanca,
grant No.~2018/14967-4. The work of DB is supported by FAPESP, grant No.~2015/20689-9 and by CNPq grant No.~309847/2018-4. The work of VM
is supported by the Spanish MINECO Ram\'on y Cajal program (RYC-2014-16022), the MECD grant \mbox{FPA2016-78645-P}, the IFT Centro de
Excelencia Severo Ochoa Program under Grant SEV-2012-0249, the EU STRONG-2020 project under the program \mbox{H2020-INFRAIA-2018-1},
grant agreement no. 824093 and the COST Action CA16201 PARTICLEFACE. DB thanks the University of Salamanca and VM thanks the University of
S\~ao Paulo in S\~ao Carlos, where parts of this work were carried out, for hospitality.
\end{acknowledgments}

\bibliography{charm2}

\end{document}